\documentclass[transmag]{IEEEtran}

\usepackage{latexsym}
\usepackage{graphicx}
\usepackage{amsfonts,amssymb,amsmath}
\usepackage{color}
\usepackage[english]{babel}
\usepackage{cite}
\usepackage{upgreek}
\usepackage{siunitx}
\usepackage{ulem}

\usepackage[colorlinks=true,linkcolor=blue,anchorcolor=blue,citecolor=red,filecolor=green,menucolor=blue,runcolor=blue,urlcolor=blue]{hyperref}

\newcommand{\ccw}{\ensuremath{\mathrm{ccw}}}
\newcommand{\cw}{\ensuremath{\mathrm{cw}}}

\newcommand{\VA}{\ensuremath{V_\mathrm{1,rms}}}
\newcommand{\VB}{\ensuremath{V_\mathrm{2,rms}}}

\newcommand{\VACW}{\ensuremath{V_\mathrm{1cw,rms}}}
\newcommand{\VBCW}{\ensuremath{V_\mathrm{2cw,rms}}}

\newcommand{\VACCW}{\ensuremath{V_\mathrm{1ccw,rms}}}
\newcommand{\VBCCW}{\ensuremath{V_\mathrm{2ccw,rms}}}
\newcommand{\wac}{\ensuremath{\omega_\mathrm{AC}}}

\def\BibTeX{{\rm B\kern-.05em{\sc i\kern-.025em b}\kern-.08em T\kern-.1667em\lower.7ex\hbox{E}\kern-.125emX}}
\begin{document}

\title{Design of electrostatic feedback for an experiment to measure $G$}

\author{Stephan Schlamminger, Leon Chao, Vincent Lee, David B. Newell, and Clive C. Speake,

\thanks{Accepted for publication in Open Journal of Instrumentation and Measurement 1 June 2022 }
\thanks{S. Schlamminger is with the National Institute of Standards and 
Technology, Gaithersburg, MD 20901 USA.}
\thanks{L. Chao is with the National Institute of Standards and 
Technology, Gaithersburg, MD 20901 USA.}
\thanks{V. Lee is with the National Institute of Standards and 
Technology, Gaithersburg, MD 20901 USA.}
\thanks{D. Newell is with the National Institute of Standards and 
Technology, Gaithersburg, MD 20901 USA.}
\thanks{C. Speake is with the University of Birmingham, Birminghan, UK.}}
\IEEEpubid{\today} 

\IEEEtitleabstractindextext{
\begin{abstract} Abstract---The torsion pendulum at the heart of the apparatus to measure the gravitational constant, $G$ at the Bureau International des Poids et Mesures (BIPM) is used to measure the gravitational torque between source and test-mass assemblies with two methods. In the Cavendish method, the pendulum moves freely. In the electrostatic-servo method, the pendulum is maintained at a constant angle by applying an electrostatic torque equal and opposite to any gravitational torque on the pendulum. The electrostatic torque is generated by a servo. This article describes the design and implementation of this servo at the National Institute of Standards and Technology. We use a digital servo loop with a Kalman filter to achieve measurement performance comparable to the one in an open loop. We show that it is possible to achieve small measurement uncertainty with an experiment that uses three electrodes for feedback control. 
\end{abstract}

}

\maketitle

\section{INTRODUCTION}

\IEEEPARstart{S}{ince} Cavendish's famous measurement, which appeared in print in 1798~\cite{Cavendish1798}, numerous researchers have attempted to measure the  Newtonian constant of gravitation, $G$. 
The adjustment of the fundamental constants performed under the auspices of the CODATA (Committee on Data of
the International Science Council) task group on fundamental constants~\cite{Tiesinga2021} gives the value of G  with a relative 1-$\sigma$ uncertainty of $2.2\times 10^{-5}$. In this article, 1-$\sigma$ refers to the standard uncertainty, i.e., a 68\,\% confidence interval. The present knowledge is a significant improvement compared to a century ago when the consensus value of $G$ could only be given with a relative uncertainty of $4.4\times 10^{-4}$ (again 1-$\sigma$). Yet, the present uncertainty of $G$ is still embarrassingly large, especially compared to other fundamental constants. It is not that the experimenters lack the ingenuity to tackle this measurement task. 
Many clever ideas (e.g.~\cite{Gundlach2000,Rosi2014}) have been applied to the experimental determination of Newton's constant. 
While individual experiments have reported relative uncertainties that are almost below  $10^{-5}$, the ensemble of measurements does not warrant such an optimistic view.

\begin{figure}
\begin{center}
    \includegraphics[width=0.95\columnwidth]{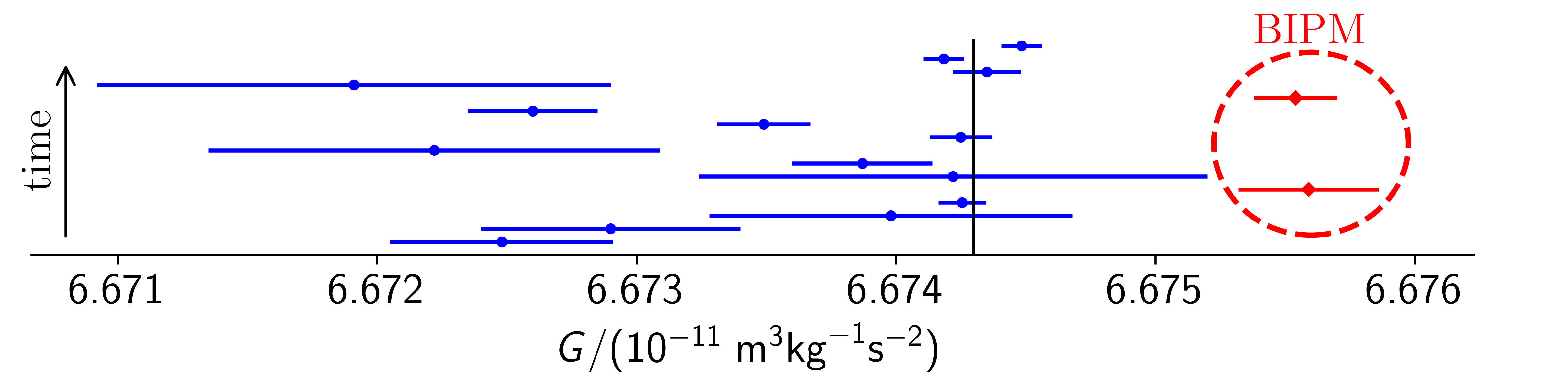}
   \caption{Measurements of $G$ over the past four decades with the most recent results on top. The two data points on the right, inside the dashed circle, were obtained by researchers at the  Bureau International des Poids et Mesures (BIPM). The torsion balance that was used in the latest result has been shipped to NIST in 2015 and is the topic of this article. All error bars are 1-$\sigma$.  The vertical line is the recommended value by CODATA (Committee on Data of the International Science Council)}
 \label{fig:G_vs_t}
 \end{center}
\end{figure}

Figure~\ref{fig:G_vs_t} shows the results of precision measurements of $G$ that were published in the last 40 years. 
Clearly, the scatter between recent results is quite large compared to the individual uncertainties. 
While the best experiments have measured $G$ with relative uncertainties close to \SI{1e-5}{}, the relative  difference between experiments can be as large as \SI{5e-4}{}. 
The reasons for such an inconsistent data set are unclear. 
Three possible explanations are widely discussed~\cite{Rothleitner2017,quinn2014newtonian}:
(1) We don't completely understand the law of gravitation,e.g.~\cite{Murata2015}, and there is a fundamental difference in these different experiments that causes their results to deviate from each other.
(2) Some or all of the experiments have at least one  subtle technical issue that is not understood completely, e.g.~\cite{Kuroda1995}, and generates a bias in  the result reported by the corresponding experiment.
(3) The experiments measure correctly, but some or all of the  assigned uncertainties are too small. If  the uncertainties were assigned correctly, the results would be consistent with the uncertainties.~\cite{Thompson2011,Merkatas2019} 

Of course, any combination of these three reasons can be the cause for the observed discrepancy.
Clearly, the first possibility is the most exciting one.
But, as metrologists, we have to do due diligence and investigate possibilities two and three.
To gain some more insight into either one of these possibilities, we decided to take one of the past experiments~\cite{Quinn2001,Quinn2013,Quinn2014a} and remeasure the gravitational constant using a different group of researchers and conduct the experiment at a separate laboratory. 
The principal investigators of the Bureau International des Poids et Mesures (BIPM)  experiment have graciously loaned the BIPM torsion balance to researchers at NIST. 

At the National Institute of Standards and Technology (NIST), the experiment was originally reconstructed with the original mechanical parts of the BIPM apparatus. 
Later on, a few hardware pieces were changed, but, the experiment is the same in function and idea.
Unlike the mechanics, the electronics and software were built almost entirely at NIST. 
This article describes the development and performance of the feedback system in detail. In the next section, we describe the physical system and introduce the theory of the electrostatic transducer. We then, describe the feedback and  show the Bode plot of the closed loop digital controller. Section 4 explains the Kalman filter that will provide an accurate estimate of the angular velocity of the pendulum. The last section shows the circuit and block diagrams of the complete system. Furthermore, a comparison of the achieved standard deviation of a measurement with the loop open and closed is given.

\section{Description of the system}

The BIPM torsion balance differs from most other torsion balances to measure $G$ in three key points. 
(1) The torsion pendulum is suspended from a ribbon, not a circular torsion fiber.
Hence, the torsion spring is comparably stiff ($\kappa=\SI{0.207}{\milli \newton \meter}$)  has a high quality factor ($Q=\SI{25000}{}$), and can carry a large load with a breaking strength  of \SI{10}{\kilo \gram}.
(2) The masses are arranged in a hexadecapole configuration.
The test masses and source masses are arranged with four-fold symmetry, see Fig.~\ref{fig:ov}. 
For such a mass arrangement, the gravitational torque is proportional to $N \propto r^4/R^5$, where $r$ and $R$ denote the radii of the  pitch circle of the test and source masses, respectively. The pitch circle is the circle that goes through the geometric center of the respective masses.
Most importantly, the torque on the pendulum depends inversely on the fifth power of the radius of the  pitch circle of the source masses and other external masses. 
The dependence on such a high power reduces the parasitic gravitational coupling of other (moving) masses in the laboratory to the torsion pendulum.
(3) One experimental apparatus can be used to measure $G$ with two different methods, also called modes: {\it Cavendish} and {\it electrostatic servo} mode.

In both cases, the torque produced by the source masses on the torsion pendulum is modulated by alternating the azimuthal orientation of the source mass assembly between two states. In one state, a clockwise (cw), and in the other, a counterclockwise (ccw) torque is produced on the test mass assembly.
The two methods differ on how the torque, or more precisely, the torque difference, is measured. 

In the Cavendish mode, the torque is inferred from the equilibrium angle of the freely moving torsion pendulum.
If an external, torque $N_{\ccw}$ acts in the counter clockwise  direction  on the pendulum, the equilibrium position $\phi_{\ccw}$ is given by $\kappa(\phi_\ccw -\phi_o)=N_\ccw$, with $\phi_o$ denoting the unknown equilibrium position in the absence of any torque. 
Hence, the torque difference is given by $N_\ccw -N_\cw=  \kappa(\phi_\ccw -\phi_\cw)$. 
Here, $N_\cw$ and $\phi_\cw$ denote the torque and equilibrium angle in the clockwise  direction. 
The only unknown quantity is the torsional stiffness $\kappa$.
It is obtained using $\kappa=4\pi^2f^2_o I_z$ from the measurement of  the torsional frequency $f_o$ and the calculated value of the moment of inertia $I_z$. 

In the electrostatic servo mode, the pendulum does not move.
Instead, a servo loop produces an electrostatic torque that holds the pendulum at a user-defined set point, $\phi_\mathrm{t}$. 
The index $\mathrm{t}$ refers to target.
Hence, the clockwise and counterclockwise gravitational torques are compensated by electrostatic torques, for example, $N_{\mathrm{el},\cw}=-N_\cw$, such that the sum of all torques on the pendulum is zero.
The gravitational torque difference is equal and opposite to the electrostatic torque difference.
The formalism to calculate the electrostatic torque is explained in greater detail below.  

Employing two methods in a single instrument provides a decisive advantage. 
If the torque differences measured in both ways are consistent within the combined measurement uncertainty, then the probability of a mistake in either (torque) measurement is small.
The physics and calculations used to obtain the two results are different, so it is implausible, but not impossible, to get an identical bias in each method. 
Likewise, the chance of introducing two canceling biases in a single method is small.  
For example, for the Cavendish method, the calculated moment of inertia of the pendulum is used; for the electrostatic method, it is not. 
Hence, agreement in the results inspires confidence in the calculation of $I_z$.

Another interesting detail is the sensitivity of the result to the angle measurement.
The result produced by the Cavendish method is proportional to the measured angle.
For the servo method, as is explained below, the calibration factor is inverse proportional to the angle reading.
Hence the average of both measured torque differences is first-order independent of a gain error in the angle readout and the difference in the measured torque difference from each method would indicate such a gain error.

Nevertheless, caution must be exercised.
It is still possible that, even when both methods produce consistent results, the results are in error. 
In case of agreement, the result should not be blindly trusted. 
A thorough investigation of possible two canceling systematic errors must be conducted. 

The measured torque difference must be divided by a mass integration constant, $\Gamma$, to obtain the gravitational constant from the measurement. 
We numerically calculate from the measured geometry  the expected torque difference $\Delta N_\mathrm{calc}$  using a reference value for $G_\mathrm{r}$. We use $G_\mathrm{r}=\SI{6.6743e-11}{\meter^3\kilo\gram ^{-1} \second^{-2}}$, although the chosen value is irrelevant.
The measured value is then obtained using
\begin{equation}
G_\mathrm{meas} = G_\mathrm{r} \frac{\Delta N_\mathrm{meas}}{\Delta N_\mathrm{calc}} 
 = \frac{\Delta N_\mathrm{meas}}{\Gamma} .
\end{equation}
The mass integration is the linchpin of this and other $G$ experiments. 
A relative error in the calculation of $\Delta N_\mathrm{calc}$ will relatively affect the result $G_\mathrm{meas}$ by the same amount independent of the employed measurement method.

\subsection{Overview of the apparatus}

\begin{figure}
\begin{center}
    \includegraphics[width=0.9\columnwidth]{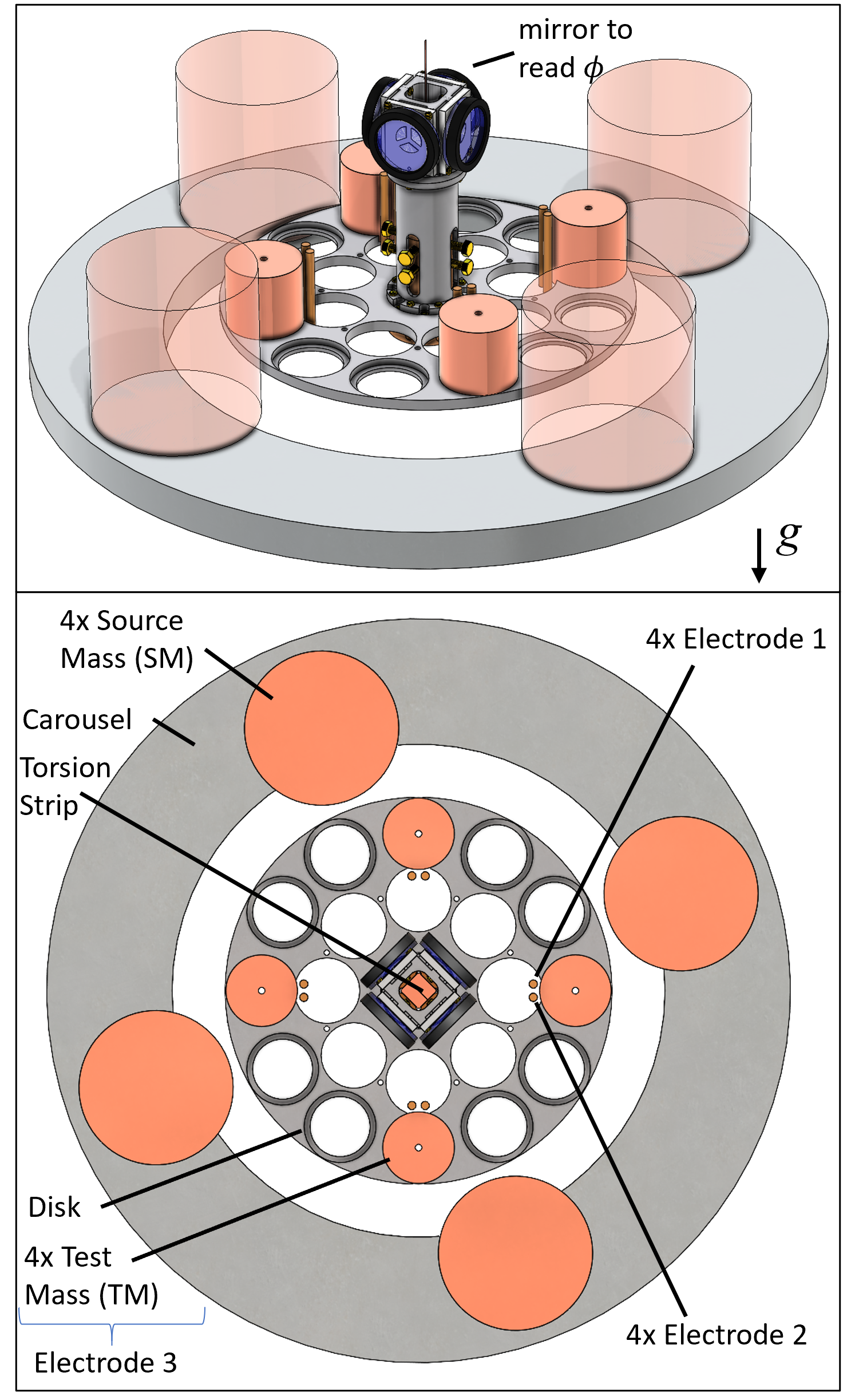}
   \caption{Top: Three-dimensional model of the torsion balance's critical components. Bottom: Bird's eye view of the system. Four cylindrical test masses on a disk suspended by a thin torsion strip inside a vacuum chamber (omitted for clarity) form a torsion balance. Outside the chamber are four cylindrical source masses on a carousel that can rotate about the strip center. The gravitational torque depends on the relative azimuthal angle between the two mass assemblies. The vacuum enclosure is electrically connected via the torsion strip to the disk and the test masses. These parts together from electrode 3. Two sets of four electrodes (labeled 1 and 2) can be used to counteract the gravitational torque. Applying a voltage to electrode  1 with respect to electrode 3 produces a counter-clockwise torque. }
 \label{fig:ov}
 \end{center}
\end{figure}

Figure~\ref{fig:ov} shows two drawings of the torsion balance and the source-mass assembly; one is a three-dimensional view, and the other is a top view. 
Two sets of electrodes labeled  1 and 2 are visible. 
Each set consists of four electrodes.
During the feedback operation, the torsion disk and test masses are connected via the torsion strip to the vacuum enclosure, which is itself grounded.
Applying a potential on the  electrodes  labeled 1, with everything else grounded, produces an electrostatic counterclockwise torque on the suspended disk. 
Torque with the opposite sign is produced when the potential is applied to the electrodes  labeled 2. 
The autocollimator reports a positive value when the pendulum rotates clockwise, as seen from the top. 

In this article, we use the mathematical convention for all angles. %
A positive angle means a counter-clockwise rotation.
For example, the angle of the torsion balance $\phi$ is the negative of the autocollimator reading.

\subsection{Electrostatic actuator}

The energy of an enclosed electrostatic system is given by
\begin{equation}
    W = \frac{1}{2}\mathbf{V}^T \mathbf{C}\mathbf{V},
    \label{eq:1}
\end{equation}
where $\mathbf{V}$ is a column vector of the applied potentials and $\mathbf{C}$ is a matrix of self and mutual capacitance, see e.g.~\cite{Smythe,Speake2005}. 
In our case, the shape of the capacitance matrix is $3\times 3$ and that of the potential  vector $3\times 1$ with  $V_1, V_2$, and $V_3$. 
The first row corresponds to electrode 1, the second to electrode 2, and the third to the pendulum plus the vacuum enclosure (electrode 3). 
The pendulum and vacuum enclosure are connected to the ground. 
While their potential is nominally 0, we allow for a contact or surface potential to exist.
This unknown potential is captured in the third row of $\mathbf{V}$.

The symmetric matrix $\mathbf{C}$, $C_{ij}=C_{ji}$, contains the self $C_{ii}$  and mutual capacitances $C_{ij}$. 
Hence, one would have to perform a total of six measurements  to construct $\mathbf{C}$.
For a closed system, only three measurements are required because the sums of each row and column are zero.
The reason is that the total charge in the system must add to zero.
The last statement is only valid if the system under consideration is entirely electrostatically enclosed.
The vacuum system in use has one window that allows the autocollimator beam to pass through. 
The system is, hence, not entirely closed. 
The deviation from a completely closed system is very small and we proceed with the assumption of a closed system.
Assuming vanishing row and column sums, the diagonal components can be obtained using
\begin{equation}
C_{ii} = -\sum_{i\ne j } C_{ij}
\end{equation}
Now, the capacitance matrix can be constructed from three  mutual capacitances $C_{12}$,$C_{13}$, and $C_{23}$.
We use a capacitance bridge to measure the cross-capacitance between two electrodes. 
The cross capacitance is the negative of the mutual capacitance~\cite{Smythe},
\begin{equation}
C_{\mathrm{c},ij} = -C_{ij}  
\end{equation}
Hence,  $\mathbf{C}$  is given by,
\begin{equation}
\mathbf{C} = \left(
\begin{array}{ccc}
C_{\mathrm{c},12}+C_{\mathrm{c},13} & -C_{\mathrm{c},12}& -C_{\mathrm{c},13} \\
-C_{\mathrm{c},12} & C_{\mathrm{c},12}+C_{\mathrm{c},23}& -C_{\mathrm{c},23} \\
-C_{\mathrm{c},13} & -C_{\mathrm{c},23}& C_{\mathrm{c},13}+C_{\mathrm{c},23} 
\end{array}
\right).
\end{equation}
It is easy to see that the entries in each row and column add to zero.
Replacing $\mathbf{C}$ in eq.~(\ref{eq:1}) with above's expression yields
\begin{eqnarray}
W = &\frac{1}{2} 
\bigg( 
C_{\mathrm{c},12} \left(V_1-V_2\right)^2 +  C_{\mathrm{c},13} \left(V_1-V_3\right)^2 
\bigg.\nonumber
\\ 
&\bigg.+  C_{\mathrm{c},23} \left(V_2-V_3\right)^2 \bigg)
\end{eqnarray}
The electrostatic torque is the negative derivative of the energy with respect to the angle of the torsion pendulum,
\begin{eqnarray}
N_{el}=-\frac{\mbox{d}W}{\mbox{d}{\phi}} =
-\frac{1}{2} 
\left( 
\frac{\mbox{d}C_{\mathrm{c},12}}{\mbox{d}\phi} \left(V_1-V_2\right)^2 +  \right.
\nonumber\\
\left.
\frac{\mbox{d}C_{\mathrm{c},13}}{\mbox{d}\phi} \left(V_1-V_3\right)^2 + 
\frac{\mbox{d}C_{\mathrm{c},23}}{\mbox{d}\phi} \left(V_2-V_3\right)^2 \right)
\end{eqnarray}
Note  the equations we define
\begin{equation}
k_{ij} :=\frac{\mbox{d}C_{\mathrm{c},ij}}{\mbox{d}\phi}
\end{equation}
The voltages applied to both control electrodes are ac voltages with a frequency of 1\;kHz.
Signals with a frequency of \SI{1}{\kilo\hertz} are used for the feedback to avoid systematic effects that could arise from a frequency mismatch between the control signals and the capacitance measurement. In this way, the control and the capacitance measurements are carried out with electrical signals of the same frequency. A frequency-dependent capacitance in the circuit would not bias the measurement result. 
The potential of the pendulum and the vacuum envelope is assumed to be a dc surface/contact potential $V_3$.
In summary,
\begin{eqnarray}
V_1 &=& \VA\sqrt{2}\sin{\left(\wac t \right)}\\
V_2 &=& \VB\sqrt{2} \sin{\left(\wac t +\alpha\right)}
\end{eqnarray}
Here, $V_\mathrm{1,rms}$ and $V_\mathrm{2,rms}$ denote the root-mean-square (rms) amplitudes of the voltages and $\alpha$ denotes the phase angle that $V_1$ lags behind $V_2$.
The two voltage values $\VA$ and $\VB$ are measured experimentally. 
The dc voltage $V_3$ is unknown. 

\begin{figure}
    \centering
    \includegraphics[width=0.9\columnwidth]{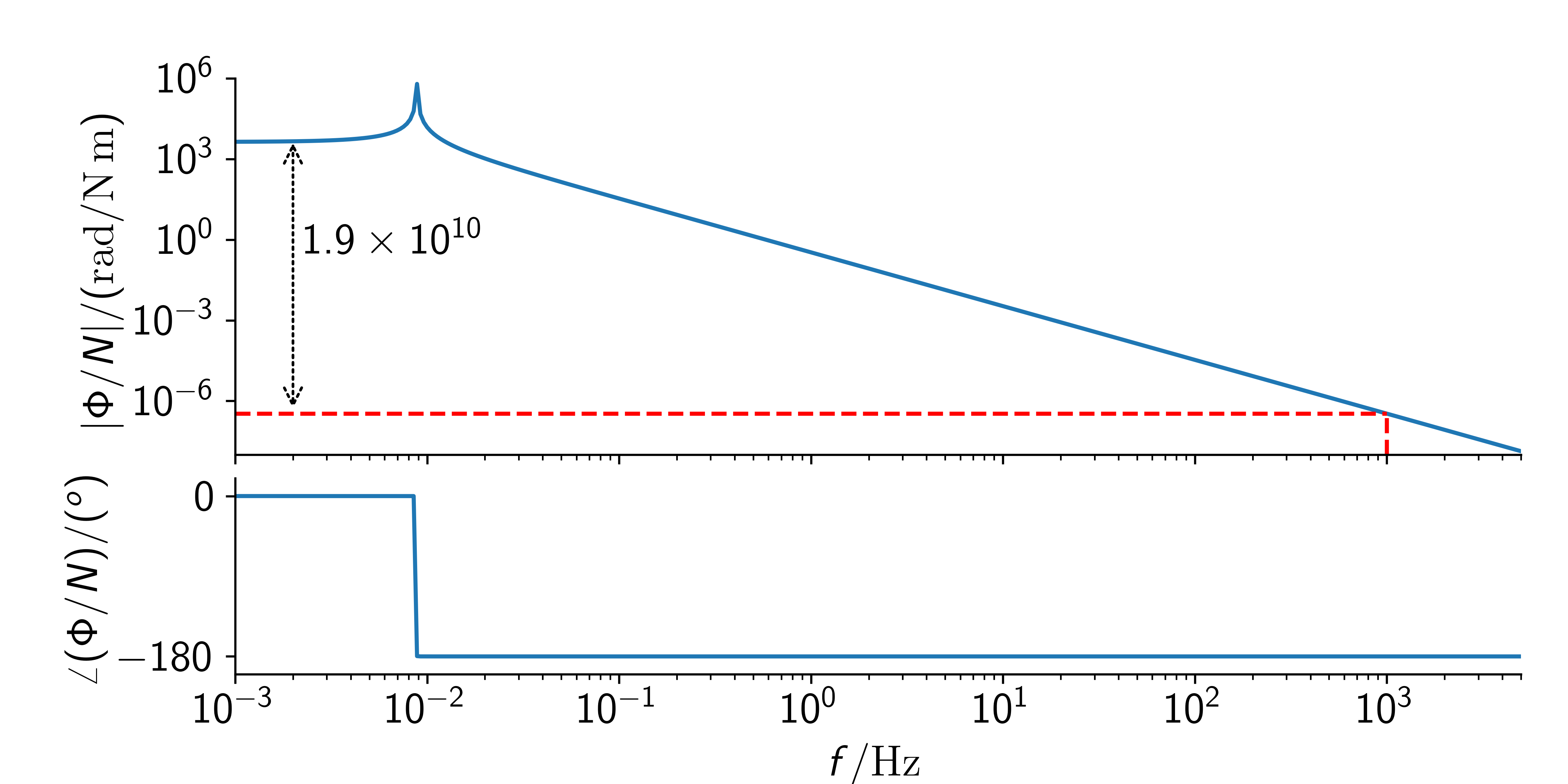}
    \caption{The continuous response function of the torsion pendulum. The top graph shows the magnitude of the response (angle divided by torque), and the bottom graph the phase of the response. The electrostatic potential varies sinusoidally with a frequency of 1\,kHz indicated by the dashed line. At this frequency, the magnitude of the response is $2\times 10^{10}$ times smaller than the response at frequencies below the resonance, as indicated by the arrow. }
    \label{fig:pend_response}
\end{figure}

The  frequency of the ac voltage (1 kHz) is not quite six orders of magnitude larger than the resonance frequency of the pendulum (1.25 mHz).
Hence, the dynamic behavior of the torsion pendulum at 1 kHz does not play any significant role,  see Fig.~\ref{fig:pend_response}, and the average torque on the pendulum is determined by  the mean squared voltages. 
The squared potential differences are integrated over one ac-period, $T_\mathrm{AC}=\SI{1}{\milli \second}$. We define for $i,j\in \{1,2,3\}$:
\begin{equation}
\langle V_{ij}^2 \rangle:= \frac{1}{T_\mathrm{AC}} \int_0^{T_\mathrm{AC}} \bigg( V_i - V_j\bigg)^2\;\mathrm{d}t
\end{equation}
For the three cases, the above definition yields
\begin{eqnarray}
\langle V_{12}^2 \rangle  &=&  \VA^2+\VB^2-2\VA\VB  \cos{\alpha},\\
\langle V_{13}^2 \rangle  &=&  \VA^2+V_3^2, \;\mbox{and}\\
\langle V_{23}^2 \rangle  &=&  \VB^2+ V_3^2.
\end{eqnarray}
With the above mean squared voltage differences the average torque on the pendulum is simply given by
\begin{equation}
    \langle N_\mathrm{el}\rangle = -\frac{k_{12}}{2} \langle V_{12}^2 \rangle    -\frac{k_{13}}{2} \langle V_{13}^2 \rangle -\frac{k_{23}}{2} \langle V_{23}^2 \rangle.
    \label{eq:torque}
\end{equation}
 The measurand is the difference in torque produced by the source masses in two different positions, labeled cw and ccw.
 For the former, the gravitational torque on the pendulum is clockwise and for the latter, it is counterclockwise. 
 The electrostatic torques are opposite the gravitational torques. 
 We use the direction of the gravitational torque as subscripts to label the corresponding voltages. 
 In the torque difference, $\Delta \langle N_\mathrm{el}\rangle=\langle N_\mathrm{grav,ccw}\rangle-\langle N_\mathrm{grav,cw}\rangle$, the dc voltage $V_3$ cancels as long as it remains constant for both measurements.
 
\begin{eqnarray}
\Delta \langle N_{el}\rangle= \frac{k_{13}}{2}  \left( \VACW^2 -\VACCW^2\right)&& \label{eq:torque:diff} \\
+\frac{k_{23}}{2}\left( \VBCW^2-\VBCCW^2\right)  &&\nonumber\\
+\frac{k_{12}}{2}\Big(\VACW^2+\VBCW^2-2\VACW\VBCW \cos\alpha\Big.\nonumber&&\\
\Big. -\VACCW^2-\VBCCW^2 +2\VACCW\VBCCW \cos\alpha\Big)&&\nonumber
\end{eqnarray}

Equations~(\ref{eq:torque}) and (\ref{eq:torque:diff}) are useful to calculate the torque on the pendulum based on measured values of $\VA$ and $\VB$. 
Since the torque is a function of these quantities squared, they introduce a non-linearity to the servo system. 

A solution is to generate the voltage amplitudes according to
\begin{eqnarray}
\VA&=&V_\mathrm{\circ,rms}+V_\mathrm{\Delta,rms}  \;\;\mbox{and}\\
\VB&=&V_\mathrm{\circ,rms}-V_\mathrm{\Delta,rms}.
\label{eq:Vtrans}
\end{eqnarray}
The voltage amplitudes are symmetric around $V_\mathrm{\circ,rms}$. One, $\VA$ is larger by $V_\mathrm{\Delta,rms}$, and the other is smaller by the same amount.
We further  assume that $k_{23}=-k_{13}$ and we make a choice of $\alpha=0$.
Since the geometry of the experiment is symmetric, the assumption is reasonable.
In reality,  their absolute values differ by less than 1\,\%. 
With the choice of the voltages above, the produced torque is almost linear in $V_\mathrm{\Delta,rms}$.  Then, Eq.~(\ref{eq:torque})  becomes, 
\begin{equation}
\langle N_{el}\rangle =  -2k_{13} V_\mathrm{\circ,rms}V_\mathrm{\Delta,rms}  -2k_{12} V_\mathrm{\Delta,rms}^2
\label{eq:torque:simple}
\end{equation}
Note the absence of $V_3$ in (\ref{eq:torque:simple}). The contact potential cancels if $k_{23}=-k_{13}$. Due to the symmetric design of the experiment, this is a reasonable assumption.
The non-linearity, i.e., the second term in equation~(\ref{eq:torque:simple}), contributes about 0.2\,\% to the total torque. 
Altogether, the assumptions and simplifications made in this paragraph are of the order percent or below. 
Such a small non-linearity is easily compensated in the closed-loop and does not affect the closed-loop system dynamics. 
To obtain a metrologically traceable result for $G$, these simplifications cannot be made. In other words, all three $k_{ij}$ must be measured.
The digital feedback system calculates the necessary electrostatic torque that needs to be applied to the pendulum. 
Solving equation~\ref{eq:torque:simple} for $V_\mathrm{\Delta,rms}$  allows the  calculation of the voltages from the desired torque. 
\subsection{The angle measurement and its noise}
The angular position of the pendulum is read with an Elcomat~HR\footnote{Certain commercial equipment, instruments, or materials (or suppliers, or software, ...) are identified in this paper to foster understanding. Such identification does not imply recommendation or endorsement by the National Institute of Standards and Technology, nor does it imply that the materials or equipment identified are necessarily the best available for the purpose.} autocollimator. 
The autocollimator reports a digital angle in the unit seconds of arc every $T_s=\SI{40}{\milli\second}$ to the host computer via a serial interface. 

\begin{figure}
\begin{center}
\includegraphics[width=0.9\columnwidth]{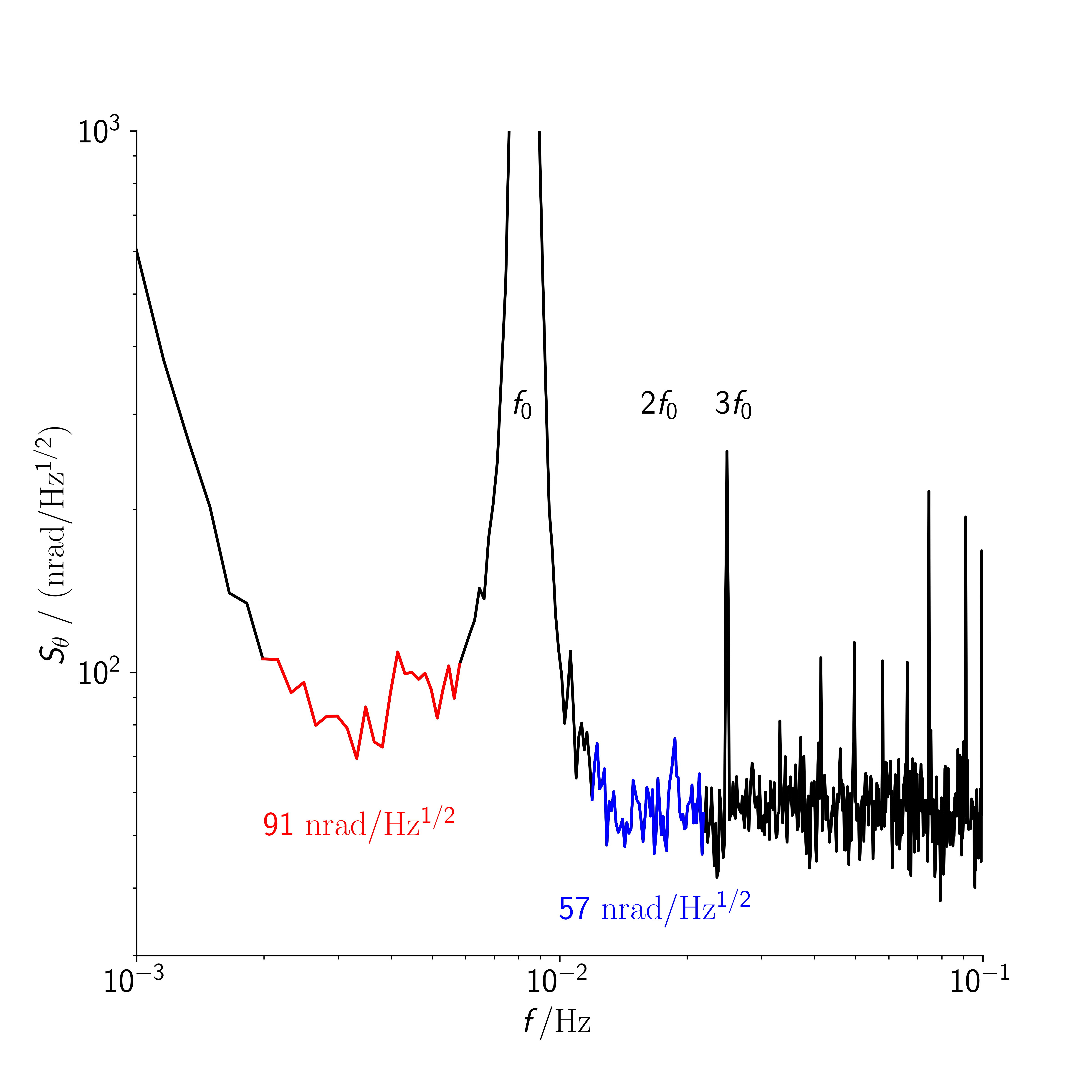}
\caption{
Amplitude spectral density of the angle. 
The pendulum amplitude was \SI{240}{\micro \radian}. The part of the spectrum that is plotted in blue determines the readout noise. The fraction of the plot shown in red is the root sum squared of the process noise and the readout noise. Due to the relatively large amplitude and nonlinearities in the autocollimator higher harmonics are visible. } 
\label{fig:noise}
\end{center}
\end{figure}

An important quantity for the servo design is the noise in the system. 
We distinguish between two different noises, read-out noise and process noise.
To establish a value for both types of noises, we observed the torsion pendulum in open loop for an extended period (\SI{8000}{\second}). 
The pendulum oscillation amplitude was  \SI{240}{\micro \radian}. 
From the measurement, an amplitude spectral density was calculated, see Figure~\ref{fig:noise}.

As is indicated in the figure, at frequencies above the resonance, the noise floor is about $S^{1/2}_{\theta}= \SI{57}{\nano\radian\per \hertz^{1/2}}$. 
The level of the noise floor is constant from the resonance to high frequencies, except of course, excess power in the harmonics. 
The data is consistent with white read-out noise. 
Digitally sampled white noise produces a spectral amplitude of
$S^{1/2}=\sqrt{2 T_\mathrm{s}}\sigma$. Hence, the uncertainty of a single autocollimator reading is $\sigma_\mathrm{autoc}=\SI{200}{\nano\radian}$.

The level of the spectral amplitude to the left side of the resonance is a consequence of at least two sources that must be added in quadrature, i.e., $\sigma^2=\sigma_\mathrm{autoc}^2+\sigma_\mathrm{N}^2$ with $\sigma_\mathrm{N} =\kappa \sigma_\theta$, where $\kappa$ is the torsional spring constant. 
Hence, we obtain $\sigma_\mathrm{N} = \SI{0.0521}{\nano \newton \meter}$ for the process noise.
This value is about $\SI{3.34e-3}{}\times \gamma_\mathrm{G}$, where $\gamma_\mathrm{G}$ is the gravitational torque produced by the source masses in the clockwise position.

For frequencies below  \SI{1}{\milli \hertz} the spectral amplitude rises proportionally to $1/f$. 
Relaxation processes in the fiber cause this rise.
It is unnecessary to include the $1/f$-noise in the model because it is like a slowly changing external torque, and the integral control of the loop will compensate for it.

\section{The control loop}
\begin{figure}
\begin{center}
\includegraphics[width=0.95\columnwidth]{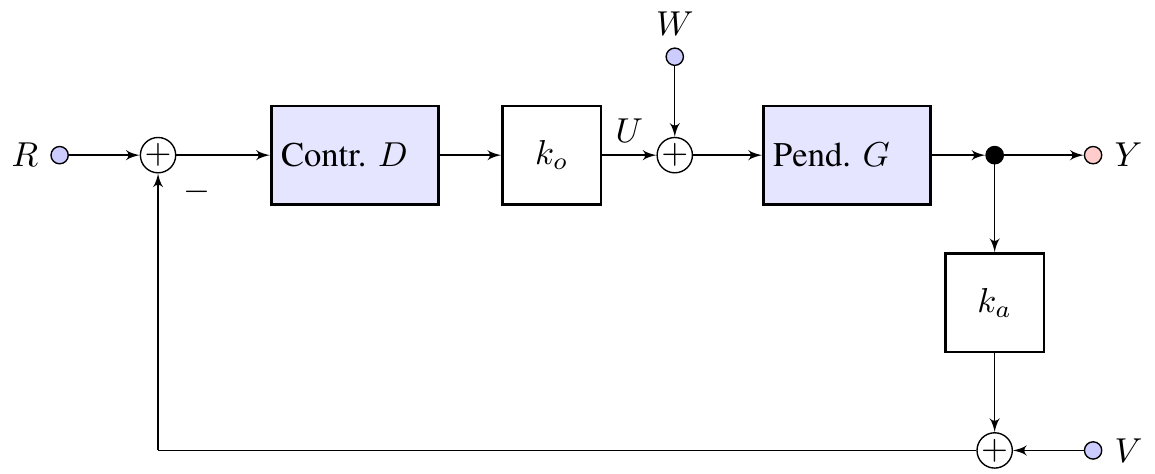}
\caption{ Block diagram of the control loop. 
The three inputs are: $R$ = reference input, $W$ = external torques on pendulum (including the gravitational torque), and  $V$= angular noise in the readout. 
The output node is the angular position of the pendulum. 
U denotes the control torque that acts on the pendulum. 
It will be a measure of W. The big boxes denote the controller (D) and the torsion pendulum(G). 
Two constants are used. One, $k_a$, converts the pendulum position in radians into seconds of arc and $k_0$ converts the torque from \SI{}{\nano \newton \meter} to \SI{}{\newton\meter}.}
 \label{fig:cd}
 \end{center}
\end{figure}

Figure~\ref{fig:cd} shows a simplified block diagram of the system. 
The digital controller is given by $D$ and the torsional oscillator by $G$. 
Two multiplicative constants are included in the loop. 
The constant  $k_\mathrm{a}=20626$ converts the output of the pendulum $Y$ from radians to seconds of arc. 
The autocollimator reports in these units.
The output of the digital controller is in $\SI{}{\nano \newton \meter}$. 
The constant $k_0=\SI{1e-9}{}$ converts that into $\SI{}{\newton \meter}$.

The system has a total of three inputs.
The desired angle set-point (reference) in seconds of arc is given by $R$.
The input $W$ is the sum of the external torques. 
These include the gravity signal, residual fiber torque, and torque noise. 
The input $V$ is the readout noise of the autocollimator in seconds of arc.

The response functions of the angle $Y$ and control signal $U$ with respect to the three inputs of reference $R$, external torque $W$, and angle noise $V$ are
\begin{eqnarray}
Y &=& \frac{G}{1+q} W -\frac{q/k_a}{1+q} \left(R - V\right)  \\
U &=& -\frac{q}{1+q} W-\frac{k_0D}{1+q} \left(R - V\right).
\end{eqnarray}
We abbreviate the open loop transfer function. with $q$, i.e.,
\begin{equation}
q=k_0 k_\mathrm{a} DG,
\end{equation}
 where $D$ and $G$ abbreviate the transfer function of the controller and the pendulum, respectively.

The electrostatic torque  $U$ is only equal and opposite to the gravitational torque $W$ when  $q$ is very high, which is the case at low frequencies. 
Ideally, $q>10^5$ 

The pendulum is a continuous system. 
Its response function is given by
\begin{equation}
    G(s) = \frac{1}{I_z(\omega_o^2+s^2)}.
\end{equation}
Here, $I_z$ denotes the moment of inertia, $\omega_o$ the angular frequency of the resonance ($\omega_o=2\pi f_o$).
Numerical values for these and other technical data  are given in Table~\ref{tab:pars} in Appendix~B.
The damping of the pendulum is so small that it can be neglected without consequences.

\begin{figure}
\begin{center}
    \includegraphics[width=0.95\columnwidth]{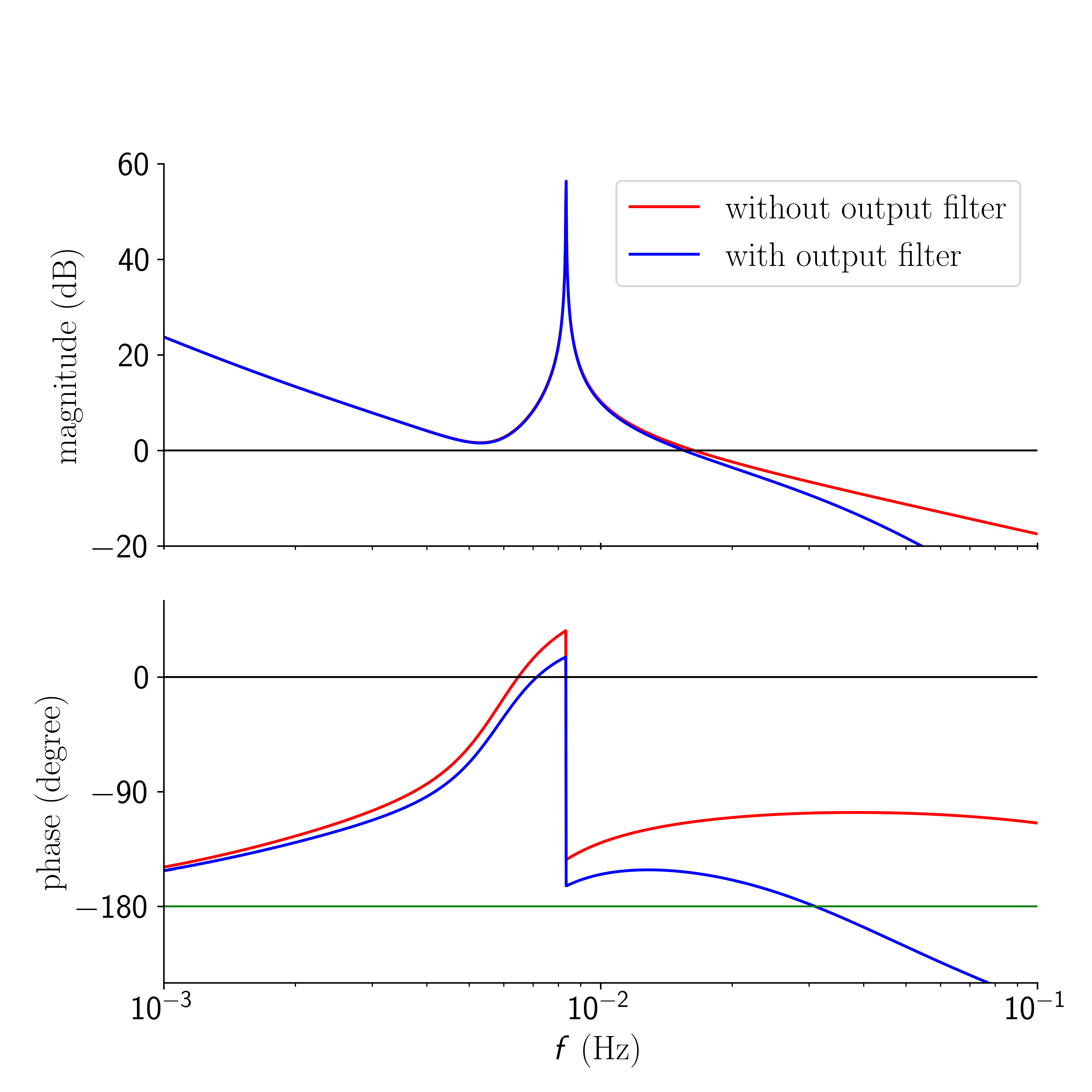}
   \caption{Bode plot of the open loop transfer function, $q$. The red trace shows $k_0k_aGD$. For the blue trace and additional output filter, $F_{\mathrm{of}}(z)$  is implemented at the output of $G$ yielding $k_0k_a G D F_{\mathrm{of}}$ for the open loop transfer function.}
 \label{fig:bode_plot}
 \end{center}
\end{figure}

The pendulum's angular excursion is sampled for every $n$-th autocollimator reading, i.e., with a loop time of $nT_s$. 
Hence, the $Z$-transform of this angle is
\begin{equation}
    G(z) =\frac{1}{I_z\omega_o^2} \frac{
    z\left(1-\cos{\left(\omega_o n T_s\right)}\right)+
    1-\cos{\left(\omega_o n T_s \right)} }
    {z^2-2 z \cos{\left(\omega_o n T_s\right)}+1  }.
\end{equation}
Using $n=15$, the known parameters of the system yield
\begin{equation}
    G(z) =\frac{2.388z+2.388 }{z^2-1.999z+1}.
\end{equation}
For the controller we assume a sum of three parts, a proportional-derivative (pd), integral (i), and double integral (ii), i.e. $D(z)=    D_\mathrm{pd}+ D_\mathrm{i}+D_\mathrm{ii}$. 
In the z-plane, the three terms are
\begin{eqnarray}
    D_\mathrm{pd}(z) &=&\frac{(k_\mathrm{p} +k_\mathrm{d}) z - k_\mathrm{d} }{z} \\
    D_{i}(z) &=&k_\mathrm{i}\frac{z}{z-1} \\
    D_{ii}(z) &=&k_\mathrm{ii}\frac{ z^2}{z^2-2z+1} 
\end{eqnarray}
The equation for $D_\mathrm{ii}(z)$ may be unfamiliar to the reader.
Appendix~\ref{app:DI} shows how it can be obtained.
It can also be obtained by squaring $D_\mathrm{i}$.
A stable feedback is obtained with $k_\mathrm{p}=1$, $k_\mathrm{d}=51$, $k_\mathrm{i}=0.03$, $k_\mathrm{ii}=0.0002$.
Figure~\ref{fig:bode_plot} shows the Bode plot of the open loop transfer function.
At high frequencies, above twice the resonant frequency, there is still too much gain in the transfer function. 
At these frequencies, there is no useful information for the $G$ measurement, but the excess voltage on the electrodes can lead to excess noise in the voltmeters that are used to read the voltage back for the precise calculation of the electrostatic torque.
This problem can be solved by filtering the output of D. 
We use a second-order filter with the transfer function
\begin{equation}
    F_{\mathrm{of}}(z) =\frac{a_0 z^2+a_1 z+a_2}{z^2+b_1z+b_2}.
\end{equation}
The numerical values of the coefficients are given in the appendix. 
The filter has a \SI{3}{\decibel} frequency of \SI{0.0314}{\hertz}. 

Finally, the reference input is also filtered with a second-order digital filter to avoid overshoot while changing the set point.
This filter has a \SI{3}{\decibel} frequency of \SI{0.0023}{\hertz}. 

\section{The Kalman observer}

The autocollimator reports a new angle reading every $T_s$. 
The controller outputs a new torque value every $n\cdot T_s$, with $n=15$. 
The reason for the lower loop frequency is so there is enough time to precisely measure $V_A$ and $V_B$.
The lower loop frequency extends the integration time of the voltmeters and ensures that the voltage on the electrodes is not changed in the middle of the measurement.
The slower update rate can be accomplished by filtering and subsequent down-sampling of the autocollimator readings.

A Kalman observer~\cite{Kalman60} offers superior performance over other filters because it includes the physics of the pendulum.
The Kalman filter yields an estimate of the state, denoted by the row vector $\mathbf{x}_k$, of the system at the time $t=k \cdot T_s$. 
We follow the description and notation given by Brown and Hwag~\cite{Brown}.

The state vector evolves according to
\begin{equation}
    \mathbf{x}_k = \mathbf{F} \mathbf{x}_{k-1} + \mathbf{G} \mathbf{u}_{k-1}+ \mathbf{w}.
    \label{eq:stateprop}
\end{equation}
Here, $\mathbf{F}$ is the state transition matrix without input.
The matrix $\mathbf{G}$ describes how the control torque $u$ influences the state. 
Lastly, the column vector $\mathbf{w}$ contains the process noise.
The readout of the state is according to 
\begin{equation}
    y_k = \mathbf{H} \mathbf{x}_{k} +v.
\end{equation}
The row vector $\mathbf{H}$ converts the state to reading to which the measurement noise $\mathbf{v}$ is added.

The state vector has three components, an unknown offset $\phi_o$, the twist of the torsion fiber $\phi_k$ and its time derivative, i.e., the angular velocity of the torsion pendulum. It is
\begin{equation}
\mathbf{x}_k = \left( 
\begin{array}{l}
\phi_o\\
\phi_k\\
\dot{\phi}_k
\end{array} \right).
\end{equation}
The state transition matrix is
\begin{equation}
\mathbf{F} = \left( 
\begin{array}{lll}
1 &0&0\\
0 & 1-\omega_0^2T_s^2&T_s/2\\
0 &-\omega_0^2T_s&1
\end{array} \right).
\end{equation}
The matrix $\mathbf{G}$ is given by
\begin{equation}
\mathbf{G} = \frac{1}{I} \left( 
\begin{array}{l}
0\\
T_s^2/2\\
T_s
\end{array} \right).
\end{equation}
Finally, the readout is determined by
\begin{equation}
\mathbf{H} =  \left( 
\begin{array}{lll}
k_\mathrm{a} & k_\mathrm{a} &0  
\end{array} \right).
\end{equation}
Note, the autocollimator measures the sum of the offset and the twist in seconds of arc, hence, $k_a$.

For the input of the control loop we use $y_k$ when $k \mod n=0$. 
Note, the matrices given above are approximations using the fact that the sample time is much shorter than the period of the pendulum, $T_s<<2\pi/\omega_0$.
The algorithm of the Kalman filter can be found in the literature~\cite{Kalman60,Brown}. 
A succinct summary is given in the appendix for completeness.

\section{The complete servo loop and its performance}

\begin{figure}
\begin{center}
\includegraphics[width=0.95\columnwidth]{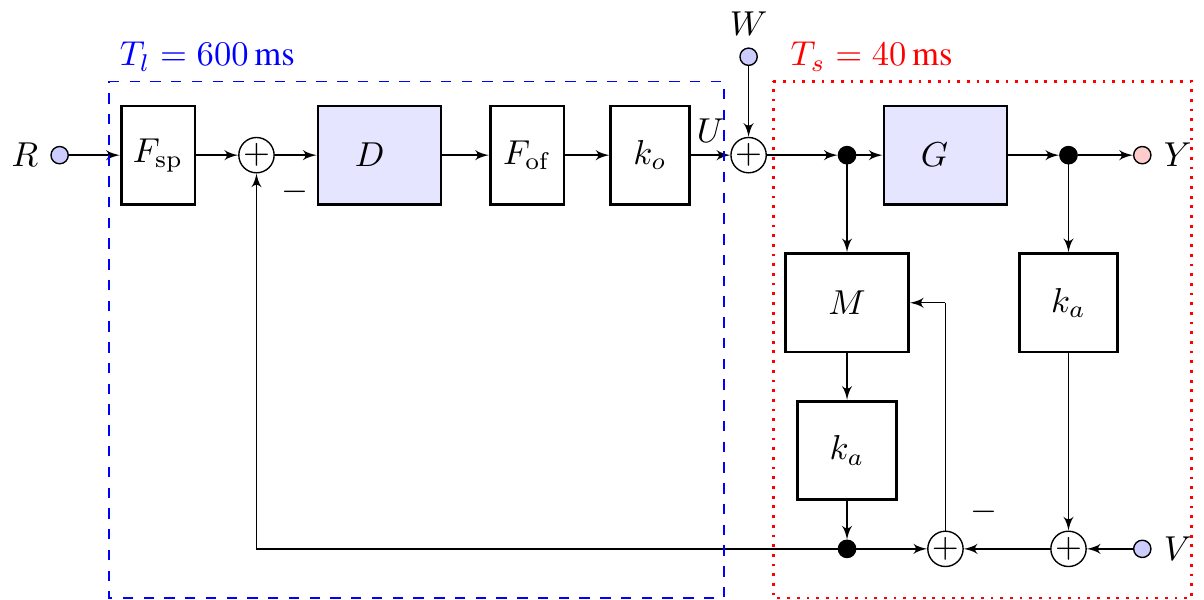}
\caption{The complete servo loop consists of two parts. They are drawn in the dashed box on the left and the dotted box on the right.
   The right box contains the angle readout with the autocollimator, $k_a$ and the Kalman filter, schematically shown as $M$. The torsion pendulum is $G$.
   The components in the right box with the exception of $G$, which is continuous, are executed every 40 ms, the readout cadence of the autocollimator.
   Every 15th calculated value from the Kalman filter is passed to the left box. It contains the controller $D$ and the output filter $F_\mathrm{of}$. The reference input $R$ can be chosen by the user. It is filtered with a set-point filter, $F_\mathrm{sp}$. }
 \label{fig:complete_loop}
 \end{center}
\end{figure}

The complete servo loop consisting of two parts is schematically shown in Fig.\ref{fig:complete_loop}. 
The part on the right, inside the dotted  box, is the Kalman filter.
The measurements and calculations are executed at a loop time of $T_s$ which is given by the readout speed of the autocollimator.
The box on the left is executed at a slower rate.
Every 15th estimate of the Kalman filter is passed to the controller $D$. 
 The requirement to downsample the autocollimator readings arises from the need to precisely measure the ac voltage at both electrodes. In order for the precision voltmeters to measure the amplitude of the  1\,kHz signal, the signal must be present with constant amplitude for some time. Downsampling the autocollimator rate by a factor of 15 is a good compromise between the performance of the loop and the voltage measurements.
The controller uses the difference between the  filtered reference input and the Kalman estimate.
The output of the controller is passed to a digital output filter ($F_\mathrm{of}$). 
The calculation and filtering are performed in units \SI{}{\nano\newton\meter}.

\begin{figure}
  \begin{center}
\includegraphics[width=0.95\columnwidth]{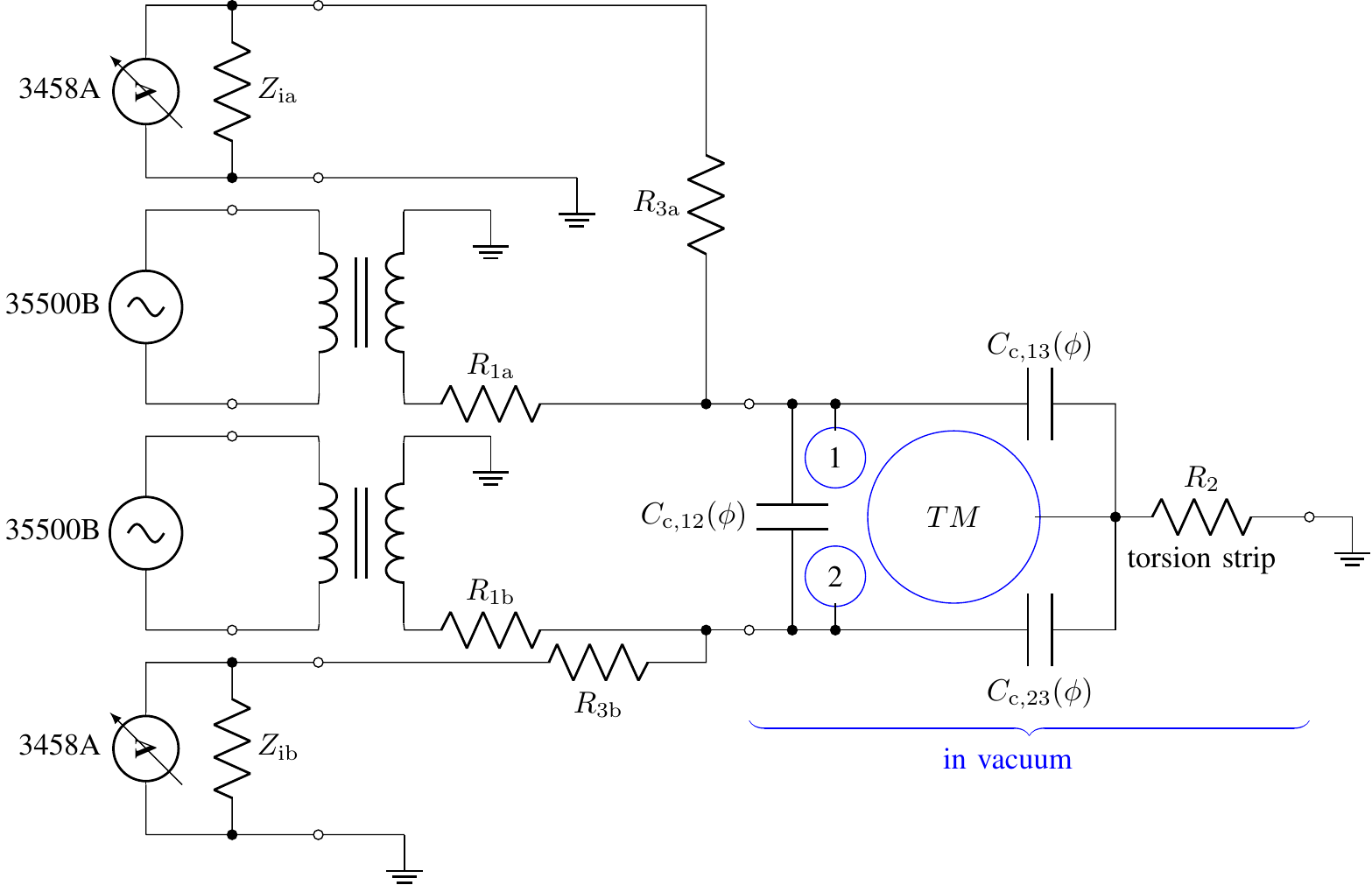}
\caption{The electrical circuit of the servo loop. On the left are the computer controller voltages sources (signal generators 35500B from Keithley) and the voltmeters (3458A from Agilent). On the right is the hardware in the vacuum enclosure, schematically one electrode 1, one electrode 2, and one test mass is shown. In reality, four of each exist. The capacitances $C_\mathrm{c,12}$,$C_\mathrm{c,13}$, and $C_\mathrm{c,23}$ depend on the angular position $\phi$ of the pendulum. The pendulum is grounded through the torsion strip, resistor $R_2$. The other impedances reflect the cable resistance and the input impedance of the voltmeter.}
 \label{fig:el_circuit}
 \end{center}
\end{figure}

Figure~\ref{fig:el_circuit} shows the electrical circuit that is used to control the pendulum and to measure the voltages on the control electrodes. 
%
The ac voltages  are generated by two signal generators (Keithley 35500B). 
They produce two phase coherent sine signals with a frequency $f=\SI{1}{\kilo\hertz}$ and a relative phase difference of $\alpha=0$, see Eq.~(\ref{eq:torque}).

The amplitude of each sine wave ranges from \SI{0}{\volt} to \SI{5}{\volt}, and is determined by the amplitude modulation input. 

The input signal to the modulation input is generated by a two-channel DAC converter in the control PC.
Each output of the signal generator is connected to a transformer that multiplies the voltage by 15.
The output of one transformer is connected to electrode $A$, and the output of the other to electrode $B$.

\subsection{Measurement of the ac voltage}

In servo control, the torque on the pendulum can only be determined by measuring the voltages on the electrodes in a metrologically traceable way. 
For that task, we employ two Agilent 3458A voltmeters.
The voltmeters are triggered by the sync output of the signal generator and measured simultaneously. 
To measure the RMS voltage of a 1 kHz waveform, we use a technique similar to the one proposed by Swerlein~\cite{Swerlein1991}. 
While Swerlein's algorithm uses the mathematical functions of the digital voltmeter to calculate the RMS voltage at its terminals, we transfer the digital readings to the PC and fit sines to the data offline. 
The advantage of our technique is that the offline data set contains the phase angle $\alpha$ between $V_1$ and $V_2$. 
Other than that we find no difference in the result to Swerlein's algorithm.

\subsection{Measurement of the capacitance gradients}

As discussed in section 2, three capacitance gradients need to be measured to infer the torque from the RMS voltage readings.
Before the capacitance measurement, the pendulum is excited to a swing amplitude of about \SI{730}{\micro \radian}. 
After that, we use relays (Ross Engineering Coorp. Model RR-1C-05-H12)  to connect an Andeen-Hagerling 2700A capacitance bridge to measure the capacitances to two of the three relevant electrical components: the clockwise pulling electrode, the counter-clockwise pulling electrodes, and the pendulum.
For each measurement, One electrode is connected to the high terminal, another to the low terminal, and the third to the ground.
Each of the three cross capacitances is measured multiple times. We describe one measurement and its analysis on the example of $C_{\mathrm{c},13}$ below.

We assume the pendulum swings in a harmonic motion, $\phi(t)=\phi_o +\phi_A \cos(\omega t)$. A series expansion of $C(\phi)$ around $\phi_o$ yields 
\begin{equation}
    C(\phi) \approx C(\phi_o) + 
    k_1 (\phi -\phi_o)
    +\frac{1}{2}  
    k_2
    \left(\phi-\phi_o\right)^2,
\end{equation}
where $k_1$ and $k_2$ denote the first and second derivative of $C$ with respect to $\phi$ at $\phi=\phi_o$. 
The capacitance gradient is smooth and an analytic equation can be found in~\cite{Quinn2014a}. For the small deflection angle of the torsion pendulum a second order series expansion is sufficient.
Hence, the measured capacitance is given by
\begin{eqnarray}
    C(t) \approx&& C(\phi_o) + \frac{k_2\phi_A^2}{4} + \nonumber \\
    &&k_1\phi_A  \cos(\omega t) + \frac{k_2}{4} \phi_A^2  \cos(2\omega t).
\end{eqnarray}
By fitting $C_o +C_A \cos(\omega t) +C_B \cos(2 \omega t)$  to the measured values $C(t)$, the capacitance gradient at $\phi_o$ can be obtained by $k_1=C_A/\phi_A$. 
The advantage of this method is that it is independent of any phase delay between the angle reading (autocollimator) and the capacitance measurement. 
Regardless of what data analysis is used, the reading of the AH bridge must be corrected for its integration time.  

\begin{figure}
\begin{center}
    \includegraphics[width=0.9\columnwidth]{
    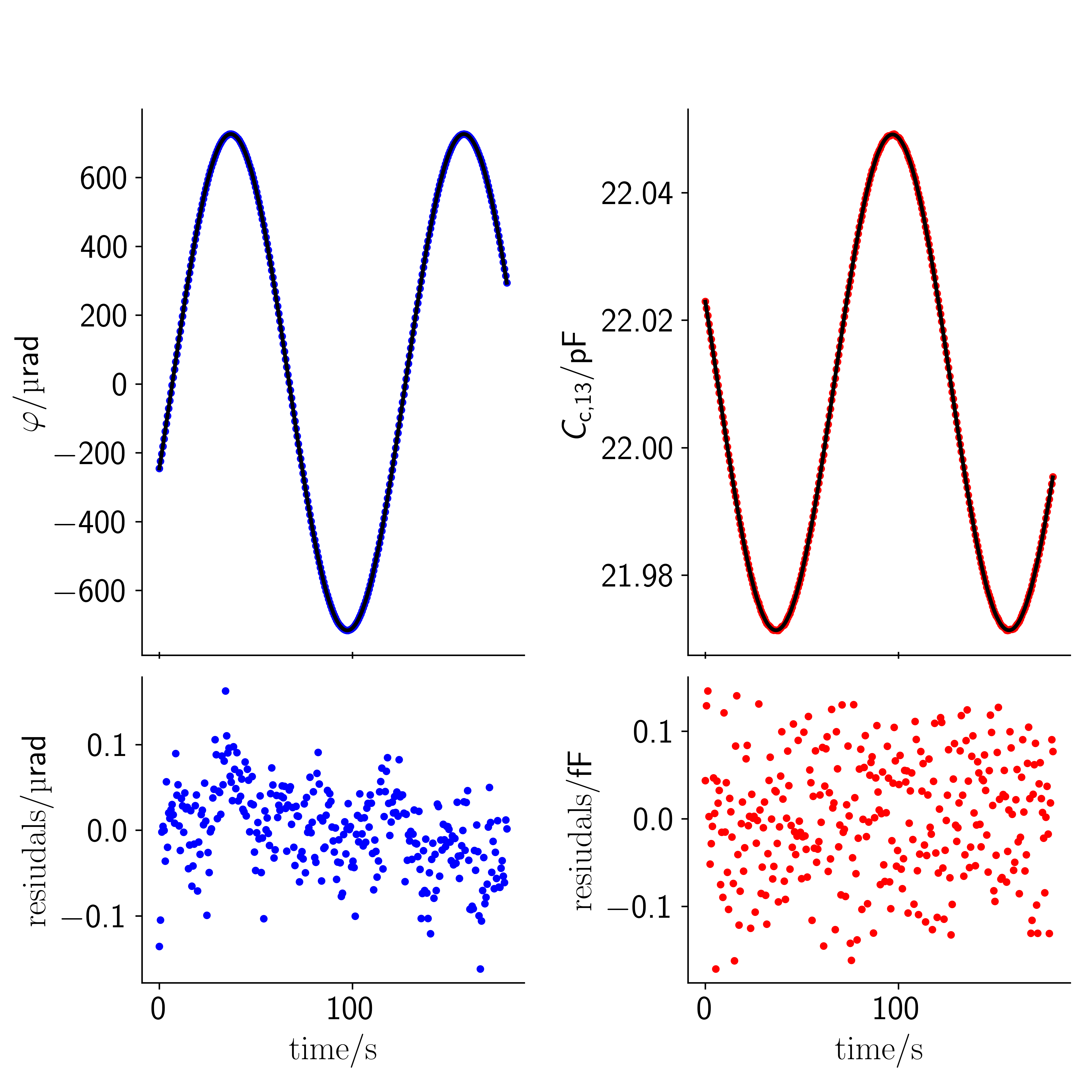}
        \caption{One data set that is used to determine the  capacitance gradient $k_{12}$.}
 \label{fig:capmeas}
 \end{center}
\end{figure}

Figure~\ref{fig:capmeas} shows an example of a capacitance gradient  measurement. 
Data is taken for a length of  1.5 pendulum periods ($\approx\SI{180}{\second}$). 
The relative uncertainty of the angle amplitude measurement is $\SI{5.8E-6}{}$, that of the capacitance measurement is $\SI{155E-6}{}$. 
 These numbers are obtained by scaling the sums of the residuals squared such that they are in agreement with the number of degrees of freedom.
Hence, about 100 measurements are required to average the relative uncertainty of the mean  down to $\SI{15.6E-6}{}$. 
The capacitance measurements are carried out periodically through the servo data measurement. Typically 15 measurements are taken every 20 hours. 
A complete $G$-data run is typically 6 days long. 

\subsection{The performance of the loop}

\begin{figure}
\begin{center}
    \includegraphics[width=0.95\columnwidth]{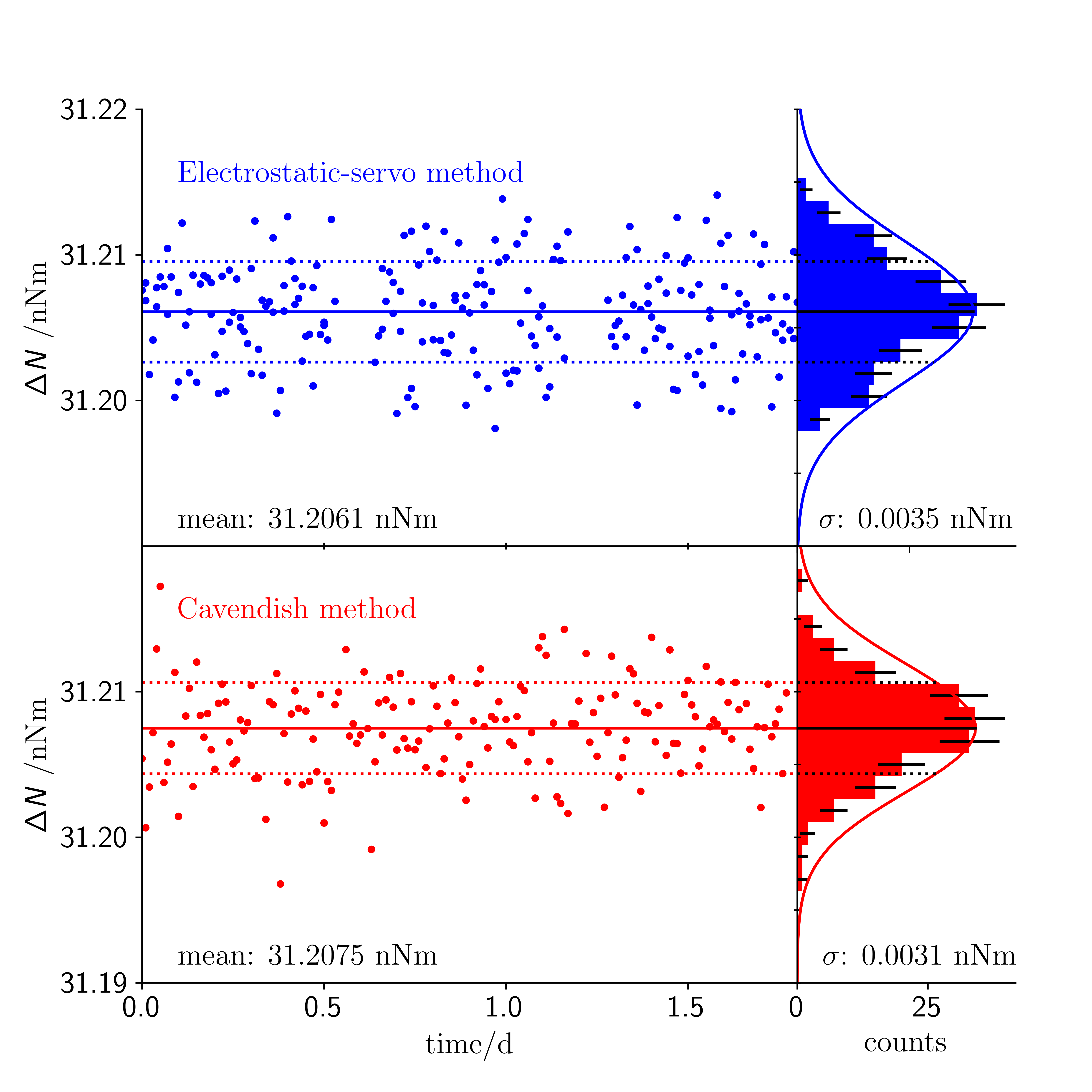}
   \caption{Comparison of the measurements of the gravitational torque difference. The top row shows the data obtained with the electrostatic-servo. The bottom row shows the data obtained with the Cavendish method. This graph shows that the servo does not signifciantly increase the noise in the measurement.}
 \label{fig:comp}
 \end{center}
\end{figure}

Figure~\ref{fig:comp} shows the measured torque difference for two data sets, each 1.75 days long. The data shown in the upper panel was measured with the electrostatic-servo method. The torque difference measured in the lower panel is obtained from the Cavendish method. In this method, the pendulum is allowed to swing freely. The torque is calculated from the change in equilibrium position times the torsion constant. The torsion constant is obtained from the calculated moment of inertia and the measured resonance frequency.

The breaks in the electrostatic servo data were used to measure the capacitance gradients, $k_{12}$, $k_{13}$, and $k_{23}$. These gradients were also measured before and after the data run.
Overall the standard deviation in the measured torque values are similar, \SI{3.5}{\pico \newton \meter} for the electrostatic-servo method and \SI{3.1}{\pico \newton \meter} for the Cavendish method.

The torque difference measured by the Cavendish method is \SI{1.4}{\pico \newton \meter} larger than the difference measured with the servo method. It is not clear why the signal is larger, but this discrepancy could point to a small error in the moment of inertia calculation.

Overall the comparison between the Cavendish and the electrostatic-servo method shows that the servo developed in this article does not introduce a significant amount of excess noise.

\section{Summary}
We have given a description of   the electrostatic servo system for the BIPM G apparatus as it is implemented at NIST.
The actuator described is different from most other electrostatic actuators used in precision measurements in two aspects. 
First, a total of three electrodes are used.
This geometry allows the production of clockwise and counterclockwise torques without storing elastic energy in the spring.
Second, two ac waveforms  each with a frequency of 1 kHz are applied to two electrodes while the third one is nominally at the ground.
The second section of this article describes the theory of such an actuator.
 In the third section, the control loop in the discrete realm is developed. 
The control loop uses a Kalman filter as a state estimator to calculate its output. 
A succinct description of the experiment-specific elements of the  Kalman filter is given in section four. 
Finally, the control loop elements are added in the fifth chapter. 
There, descriptions of all necessary measurements and the performance of the control loop are given. 

Controlling a harmonic oscillator is a common occurrence in physics and engineering. 
We hope the description of the control loop for the big G experiment can be easily transferred to other experiments, and the reader gains valuable insight from our control loop design.

\appendices

\section{Technical Data}
\label{sec:pars}
Table~\ref{tab:pars} provides a summary of the most important technical parameters of the system. 

\begin{table}[h!]
\caption{Important technical data.\label{tab:pars}}
\begin{tabular}{ l c  S[table-format=12.2] l }
sample time & $T_s$ & 0.04&\SI{}{\second}\\
spring constant & $\kappa$ & 0.207&\SI{}{\milli \newton \meter}\\
moment of Inertia & $I_z$ & 0.075
&\SI{}{\kilo\gram\meter^2}  \\ 
quality factor & $Q$ 25000\\
resonance frequency  &$f_o$& 8.28&\SI{}{\milli\hertz} \\
gravitational torque & $\Gamma_\mathrm{G}$ &  \pm15.586  & \SI{}{\nano \newton \meter}\\
cap. gradient el. 13 
&$\mbox{d}C_{\mathrm{c},13}/\mbox{d}\theta$& 
-54.235&
\SI{}{\pico \farad \per \radian}\\ 
cap. gradient el. 23 
&$\mbox{d}C_{\mathrm{c},23}/\mbox{d}\theta$& 
+55.650&
\SI{}{\pico \farad \per \radian}\\ 
cap. gradient 12 &$\mbox{d}C_{\mathrm{c},12}/\mbox{d}\theta$& 
1.296& \SI{}{\pico \farad \per \radian}\\  
transformer ratio & $\eta$ & 18.2 \\
control factor  & $k_\mathrm{el}$ & 45.763&\SI{}{\nano \newton \meter \per \volt}\\
autocollimator noise & $\sigma_\mathrm{autoc}$ & 200 &\SI{}{\nano \radian}\\
torque  noise & $\sigma_\mathrm{N}$ & \SI{0.0521}{} &\SI{}{\nano\newton \meter}\\
noise/signal & $\sigma_\mathrm{N}/\Gamma_\mathrm{G}$ & \SI{3.341e-3}{} &\\ 
\end{tabular}
\end{table}

Two digital filters are used in the control loop. The filter coefficients of the output filter are given in Table~\ref{tab:opf}. Likewise, the coefficients of the set point filter are given in Table~\ref{tab:spf}. For both filters the equation is
\begin{equation}
    F_{\mathrm{of}}(z) =\frac{a_0 z^2+a_1 z+a_2}{z^2+b_1z+b_2}
\end{equation}
with the filter coefficients printed in the corresponding tables.

\begin{table}[h!]
\caption{Filter coefficients of the output filter.\label{tab:opf}}
\begin{tabular}{ l S[table-format=8.5]}
$a_0$ & 0.00502\\
$a_1$ & 0.01004\\
$a_2$ & 0.00502\\
$b_1$ & -1.7497\\
$b_2$ &  0.7698\\
\end{tabular}
\end{table}

\begin{table}[h!]
\caption{Filter coefficients of the set point filter filter.\label{tab:spf}}
\begin{tabular}{ l S[table-format=8.5e4]}
$a_0$ & 3.16544e-5\\
$a_1$ & 6.33088e-5\\
$a_2$ & 3.16544e-5\\
$b_1$ & -1.98047\\
$b_2$ &  0.98061\\
\end{tabular}
\end{table}

\section{Calculations for the Kalman filter}
\label{app:Kalman}

At the beginning of the process we calculate the prediction of the state vector and its covariance matrix. The prediction of the state vector is
\begin{equation}
\mathbf{x}_{p,k}= \mathbf{F} \mathbf{x}_{k-1}  + \mathbf{G} \mathbf{u}_{k-1}.
\end{equation}
The prediction of the covariance matrix is
\begin{equation}
\mathbf{P}_{p,k}= \mathbf{F} \mathbf{P}_{k-1}  + \mathbf{Q},
\end{equation}
where $\mathbf{Q}$ is the process noise covariance matrix. With that the Kalman gain can be calculated. It is 
\begin{equation}
    \mathbf{K}_k =  \mathbf{P}_{p,k} \mathbf{H}^T \left(
    \mathbf{H}\mathbf{P}_{p,k}\mathbf{H}^T+\mathbf{R} \right)^{-1},
\end{equation}
where $\mathbf{R}$ is the measurement noise covariance, i.e., $v^2$.
The Kalman gain $\mathbf{K}_k$ determines what linear combination of the predicted internal state and observation is used for the calculation of the new internal state. The new internal state is
\begin{equation}
    \mathbf{x}_k =  \mathbf{x}_{p,k} + \mathbf{K}_k (y_{k} - \mathbf{H} \mathbf{x}_{p,k}).
\end{equation}
In the same way, the covariance matrix of the internal state is propagated from the predicted covariance matrix using
\begin{equation}
    \mathbf{P}_k =  \left(\mathbf{I}-\mathbf{K}_k  \mathbf{H} \right) \mathbf{P}_{p,k}.
\end{equation}

\section*{Discrete double integrator}
\label{app:DI}
We are looking for $y$ that is the double integral of $x$ with respect to time,
\begin{equation}
y(t) = k_{ii} \iint_0^t x(t')\mbox{d}t' 
\end{equation}
Hence,
\begin{equation}
 y''=  c_{ii} x.
\end{equation}
Discretizing the double differentiation yields
\begin{equation}
 y[n]-2y[n-1]+y[n-2] = c_{ii} x[n]
\end{equation}
This equation can be written in the $z$ domain as
\begin{equation}
 Y-2 z^{-1} Y+z^{-2} Y = c_{ii} X.
\end{equation}
Hence,
\begin{equation}
\frac{Y}{X} =  \frac{ c_{ii} }{1-2z^{-1}+z^{-2}}= 
\frac{ c_{ii}z^2 }{z^2-2z+1}
\end{equation}


\begin{thebibliography}{10}
\providecommand{\url}[1]{#1}
\csname url@samestyle\endcsname
\providecommand{\newblock}{\relax}
\providecommand{\bibinfo}[2]{#2}
\providecommand{\BIBentrySTDinterwordspacing}{\spaceskip=0pt\relax}
\providecommand{\BIBentryALTinterwordstretchfactor}{4}
\providecommand{\BIBentryALTinterwordspacing}{\spaceskip=\fontdimen2\font plus
\BIBentryALTinterwordstretchfactor\fontdimen3\font minus
  \fontdimen4\font\relax}
\providecommand{\BIBforeignlanguage}[2]{{%
\expandafter\ifx\csname l@#1\endcsname\relax
\typeout{** WARNING: IEEEtran.bst: No hyphenation pattern has been}%
\typeout{** loaded for the language `#1'. Using the pattern for}%
\typeout{** the default language instead.}%
\else
\language=\csname l@#1\endcsname
\fi
#2}}
\providecommand{\BIBdecl}{\relax}
\BIBdecl

\bibitem{Cavendish1798}
\BIBentryALTinterwordspacing
H.~Cavendish, ``Experiments to determine the density of the earth,''
  \emph{Philos. Trans. Roy. Soc. London}, vol.~88, pp. 469--526, 1798.
  [Online]. Available: \url{http://www.jstor.org/stable/106988}
\BIBentrySTDinterwordspacing

\bibitem{Tiesinga2021}
\BIBentryALTinterwordspacing
E.~Tiesinga, P.~J. Mohr, D.~B. Newell, and B.~N. Taylor, ``Codata recommended
  values of the fundamental physical constants: 2018,'' \emph{Rev. Mod. Phys.},
  vol.~93, p. 025010, Jun 2021. [Online]. Available:
  \url{https://link.aps.org/doi/10.1103/RevModPhys.93.025010}
\BIBentrySTDinterwordspacing

\bibitem{Gundlach2000}
\BIBentryALTinterwordspacing
J.~H. Gundlach and S.~M. Merkowitz, ``Measurement of newton's constant using a
  torsion balance with angular acceleration feedback,'' \emph{Phys. Rev.
  Lett.}, vol.~85, pp. 2869--2872, Oct 2000. [Online]. Available:
  \url{http://link.aps.org/doi/10.1103/PhysRevLett.85.2869}
\BIBentrySTDinterwordspacing

\bibitem{Rosi2014}
G.~Rosi, F.~Sorrentino, L.~Cacciapuoti, M.~Prevedelli, and G.~M. Tino,
  ``Precision measurement of the {N}ewtonian gravitational constant using cold
  atoms,'' \emph{Nature}, vol. 510, no. 7506, pp. 518--521, jun 2014.

\bibitem{Rothleitner2017}
\BIBentryALTinterwordspacing
C.~Rothleitner and S.~Schlamminger, ``Invited review article: Measurements of
  the newtonian constant of gravitation, g,'' \emph{Review of Scientific
  Instruments}, vol.~88, no.~11, p. 111101, 2017. [Online]. Available:
  \url{https://doi.org/10.1063/1.4994619}
\BIBentrySTDinterwordspacing

\bibitem{quinn2014newtonian}
T.~Quinn and C.~Speake, ``The newtonian constant of gravitation--a constant too
  difficult to measure? {A}n introduction,'' \emph{Phil. trans. R. Soc. A},
  vol. 372, no. 2026, 2014.

\bibitem{Murata2015}
J.~Murata and S.~Tanaka, ``A review of short-range gravity experiments in the
  {LHC} era,'' \emph{Class. Quantum Grav.}, vol.~32, p. 033001, 2015.

\bibitem{Kuroda1995}
\BIBentryALTinterwordspacing
K.~Kuroda, ``Does the time-of-swing method give a correct value of the
  {N}ewtonian gravitational constant?'' \emph{Phys. Rev. Lett.}, vol.~75,
  no.~15, pp. 2796--2798, Oct. 1995. [Online]. Available:
  \url{http://link.aps.org/doi/10.1103/PhysRevLett.75.2796}
\BIBentrySTDinterwordspacing

\bibitem{Thompson2011}
M.~Thompson and S.~L.~R. Ellison, ``Dark uncertainty,'' \emph{Accreditation and
  Quality Assurance}, vol.~16, no.~10, p. 483, 2011.

\bibitem{Merkatas2019}
\BIBentryALTinterwordspacing
C.~Merkatas, B.~Toman, A.~Possolo, and S.~S., ``Shades of dark uncertainty and
  consensus value for the newtonian constant of gravitation,''
  \emph{Metrologia}, vol.~56, no.~5, p. 054001, aug 2019. [Online]. Available:
  \url{https://doi.org/10.1088\%2F1681-7575\%2Fab3365}
\BIBentrySTDinterwordspacing

\bibitem{Quinn2001}
T.~J. Quinn, C.~Speake, S.~J. Richman, R.~S. Davis, and A.~Picard, ``A new
  determination of {$G$} using two methods,'' \emph{Phys. Rev. Lett.}, 2001.

\bibitem{Quinn2013}
\BIBentryALTinterwordspacing
T.~Quinn, H.~Parks, C.~Speake, and R.~Davis, ``{Improved Determination of {$G$}
  Using Two Methods},'' \emph{Phys. Rev. Lett.}, vol. 111, p. 101102, Sep 2013.
  [Online]. Available:
  \url{http://link.aps.org/doi/10.1103/PhysRevLett.111.101102}
\BIBentrySTDinterwordspacing

\bibitem{Quinn2014a}
T.~Quinn, C.~C. Speake, H.~Parks, and D.~R., ``The {BIPM} measurements of the
  newtonian constant of gravitation, {$G$},'' \emph{Phil. Trans. R. Soc. A},
  vol. 372, p. 20140032, 2014.

\bibitem{Smythe}
W.~Smythe, \emph{Static and Dynamic Electricity}.\hskip 1em plus 0.5em minus
  0.4em\relax New York and London: McGraw-Hill Book Company, 1939.

\bibitem{Speake2005}
C.~C. Speake, ``Newton's constant and the twenty-first century laboratory,''
  \emph{Phil. Trans. R. Soc. A}, vol. 363, no. 1834, pp. 2265--2287, 2005.

\bibitem{Kalman60}
R.~E. Kalman, ``A new approach to linear filtering and prediction problems,''
  \emph{Transactions of the ASME--Journal of Basic Engineering}, vol.~82, no.
  Series D, pp. 35--45, 1960.

\bibitem{Brown}
R.~Brown and P.~Hwag, \emph{Introduction to Random Signals and applied Kalman
  Filtering}.\hskip 1em plus 0.5em minus 0.4em\relax New York: John Wiley \&
  Sons, Inc., 1992.

\bibitem{Swerlein1991}
L.~Swerlein, ``A 10 ppm accurate digital ac measurement algorithm,''
  Hewlwett-Palcard internal publication, Tech. Rep., 1991.

\end{thebibliography}
\end{document}